# Robustness of Floquet topological phase at room temperature: a first-principles dynamics study

Ruiyi Zhou[1] and Yosuke Kanai[1,2*]

1. Department of Chemistry, University of North Carolina at Chapel Hill, Chapel Hill, North Carolina 27514, USA
2. Department of Physics and Astronomy, University of North Carolina at Chapel Hill, Chapel Hill, North Carolina 27514, USA

Corresponding author email: ykanai@unc.edu

**Abstract**

Nonadiabatic Thouless pumping of electrons is studied within the framework of topological Floquet engineering, particularly focused on how atomic lattice dynamics affect the emergent Floquet topological phase in *trans*-polyacetylene under the driving electric field. As similarly done in the earlier work [Zhou and Kanai, J. Phys. Chem. Lett., 12, 4496 (2021)], the real-time time-dependent density functional theory and Ehrenfest dynamics simulations were used to investigate the extent to which the number of pumped charges remains equal to the topological invariant, the winding number, when the temperature effect of ions and the dynamical coupling of electrons and ions are taken into account. Our theoretical work shows that the Floquet topological phase remains intact but the condition on the driving field necessary for observing the topological phase becomes more limiting.

**Introduction**

An earliest realization of how topological characteristics of Hamiltonian govern certain dynamical properties came from Thouless when he discussed the quantized pumping phenomenon in 1983[1]. Studying one-dimensional quantum-mechanical particle transport in a slow varying potential, he showed that the quantization of the number of transported quantum particles derives from the topology of the underlying Hamiltonian. Under the adiabatic evolution (i.e. instantaneous eigenstates of Hamiltonian), the particle current in a one-dimensional system is given by the topological quantity called winding number when the Hamiltonian is periodic in time. For topological materials, the winding number can take a non-zero integer value while it is zero for normal/trivial insulators. In recent years, Thouless pumping has been demonstrated experimentally in various systems[2-5] including an ultracold Fermi gas[6] and ultracold atoms in optical lattice[7]. Various theoretical studies[8-11] have employed model Hamiltonians such as the Rice-Mele model[12] and the description of topological pumping has often assumed a complete adiabaticity of the Hamiltonian evolution. At the same time, studies of nonadiabatic effects on Thouless pumping have appeared in the literature[13-15]. A particularly notable development in recent years is the idea of the so-called topological Floquet engineering, in which one uses a time-periodic external field to induce a topological phase in the driven system that is otherwise a trivial insulator[16-17]. Briefly, in a Floquet system, the time-dependent Hamiltonian satisfies $\widehat{H}(t+T) = \widehat{H}(t)$ and an effective time-independent Hamiltonian can be defined from the time evolution operator. One can then apply the topological description analogously to this effective Hamiltonian. Under certain conditions on the driving field, the Floquet topological phase, in which the winding number is a non-zero integer, can emerge.

There is an increasing effort to develop a molecular-level understanding of topological materials from the perspective of chemistry[18-21], which would open up the field for further exploration with a more intuitive viewpoint based on the arrangement of atoms. In our previous work[22], we have demonstrated nonadiabatic Thouless pumping of electrons in *trans*-polyacetylene under a driving electric field using first-principles theory in the framework of Floquet engineering. By exploiting the gauge invariance of the quantum dynamics in terms of the minimal particle–hole excitations[23], the topological pumping of electrons can also be understood as a cyclic transition among the bonding and antibonding orbitals. Having connected the topological invariant to the chemical bonding concepts[22], we further demonstrated molecular-level control on the emergence of the Floquet topological phase, presenting an exciting possibilities of the intuitive molecular-level engineering for such an exotic topological phase.[24] At the same time, the feasibility of observing Floquet topological pumping at the room temperature still remains an practical open question. In this work, we investigate the interplay between the electronic current and the lattice dynamics of atoms on the Floquet topological phase using Ehrenfest dynamics simulation based on real-time time-dependent density functional theory (RT-TDDFT). Additionally, we examine the impact of thermal fluctuations in the atomic lattice on the Floquet topological phase by performing RT-TDDFT calculation with first-principles molecular dynamics simulation.

**Computational Details**

We examined several exchange-correlation (XC) functionals by calculating the equilibrium geometry of trans-polyacetylene as shown in Table 1. Popular XC approximations of GGA:PBE[25], meta-GGA:SCAN[26], and hybrid-GGA:PBE0[27] do not yield the Peierls distortion[28]. This failure is primarily due to self-interaction errors that lead to an over-delocalization of electrons, incorrectly favoring an undistorted structure. In contrast, the SCAN0 approximation[29], which incorporates 25% exact exchange within a meta-GGA framework, correctly predicts the Peierls distortion, yielding the bond lengths in good agreement with experimental data[30]. The accurate representation of bond length alternation is necessary here because these structural details govern the electronic structure and, consequently, the material's electrical conductivity properties. Inaccuracies in bond lengths, as seen with the other XC approximations, would indirectly lead to erroneous predictions of these key characteristics. Therefore, we employ SCAN0 XC functional for examining lattice dynamical effects in our study.[31] To simulate trans-polyacetylene at room temperature (300K), we performed first-principles molecular dynamics (FPMD) simulations using SCAN0 XC functional with the FHI-aims code[32-33]. We employ the tier 2 numeric atom-centered orbital (NAO) basis set[33]. The FPMD simulations were carried out in the NVT ensemble at 300K for 4 picoseconds, employing a Nose-Hoover thermostat[34-35] with the thermostat frequency of 1600 $cm^{-1}$.[36] The FPMD time step of 0.5 femtoseconds was used.

In order to study the Floquet topological phase by simulating nonequilibrium electron dynamics, we perform real-time time-dependent density functional theory (RT-TDDFT)[37-39] and Ehrenfest dynamics[23] based on RT-TDDFT. The mean-field approximation inherent in Ehrenfest dynamics for coupled electron-ion systems imposes limitations on the description of certain energy exchange phenomena, such as Joule heating[40-41]. However, the characteristic timescale of the Floquet topological pump under investigation (approximately several femtoseconds) is significantly shorter than the time required to reach the detailed balance condition (typically hundreds of femtoseconds to picoseconds or longer). Therefore, we posit that Ehrenfest dynamics provides a sufficient framework for describing the pertinent physics in this study. For Ehrenfest dynamics simulations, we use the Qb@ll branch[42-46] of the Qbox code[47] within the plane-wave pseudopotential (PW-PP) formalism[48]. All atoms were represented by Hamann-Schluter-Chiang-Vanderbilt (HSCV) norm-conserving pseudopotentials[49-50], with the plane-wave cutoff energy of 40 Ry for the Kohn-Sham (KS) orbitals. For integrating the time-dependent Kohn-Sham equation, we employed the enforced time-reversal symmetry (ETRS) integrator[51] in the maximum localized Wannier functions (MLWF) gauge[52], and we used 0.1 a.u. integration step size. For Ehrenfest dynamics simulation, atomic positions were also updated at the time interval of 0.1 a.u. for each integration step. The electric field $\boldsymbol{E}(t) = \boldsymbol{A} \sin\left(\frac{2\pi}{T}t\right)$ was applied as the driving field, and we considered the time periodicity $T$ of 50~200 a.u. and the field amplitude $|\boldsymbol{A}|$ range of $1.0$~$6.0 \times 10^{-3}$ a.u. with the uniform sampling intervals of 25 a.u. and $1.0 \times 10^{-3}$ a.u., respectively. Note that the time-dependent electric field was applied in the length gauge using the scalar potential in the KS Hamiltonian[46, 52]. For all simulations, a 55-atom simulation supercell, consisting of 11 C-C monomer units along the x-axis, was employed with the periodic boundary conditions ($51.32 \times 15.0 \times 15.0$ Bohr), and the Γ-point approximation for Brillouin zone integration was adopted as done in previous work[22, 24]. The comparison between the FHI-aims and Qb@ll codes is provided in Table S1 in the Supporting Information.

| XC functional | C-C bond length (Angstrom) | C=C bond length (Angstrom) | Peierls Distortion |
|---|---|---|---|
| PBE | 1.396 | 1.396 | No |
| SCAN | 1.387 | 1.387 | No |
| Hartree Fock | 1.332 | 1.460 | Yes |
| PBE0 | 1.393 | 1.393 | No |
| SCAN0 | 1.357 | 1.419 | Yes |
| Experiment | 1.36 | 1.44 | Yes |

**Table 1**: Comparison of carbon-carbon bond lengths of the optimized equilibrium structures using several exchange-correlation (XC) functionals, together with the experimental values reported in Ref.[30].

## Results and Discussion

We study two different aspects of the influence of atomic lattice dynamics on the nonadiabatic Thouless pumping of electrons. We examine how the quantum dynamics of electrons can induce the atomic lattice dynamics and potentially degrade the intricated nature of the topological pumping even at 0K. We do so by taking into account the nonadiabatic coupling between electrons and classical nuclei in the Ehrenfest dynamics simulation. In this simulation, the initial temperature of 0K is used for the lattice. Secondly, we examine the thermal fluctuations of the atomic lattice on the quantum dynamics of electrons. Because of the vastly different relevant timescale differences (i.e. femto-seconds for the pumping dynamics vs. picoseconds for the lattice dynamics), we simple take snapshots from FPMD simulations at 300K and perform RT-TDDFT on those static structures.

### Coupling between lattice dynamics and electron transport

As shown in our previous work[22, 24], the (time-) integrated current, $Q(T)$, can be calculated via time-dependent maximally-localized Wannier functions (MLWFs), $w_i(r,t)$, as[22]

$$Q(T) = L^{-1} \sum_{i}^{Occ.} [\langle w_i(T)|\hat{r}|w_i(T)\rangle - \langle w_i(t=0)|\hat{r}|w_i(t=0)\rangle] \quad (1)$$

where the position operator $\hat{r}$ is defined according to the formula given by Resta for extended periodic systems[53] and $L$ is the lattice length of the unit cell. In short, the integrated current $Q(T)$ is equal to the number of the geometric centers of the MLWFs (i.e. Wannier centers) transported over time $T$.[54-55] In the context of Floquet theory, the initial state must return to the same Hilbert subspace of the Floquet operator after a driving period. The Kohn-Sham Hamiltonian depends on the time-dependent electron density even when the adiabatic approximation is adapted for the exchange-correlation potential[56-57]. Therefore, the Hamiltonian is time-periodic (i.e. Floquet theory $\widehat{H}(t+T) = \widehat{H}(t)$ is applicable) only if the initial electron density is recovered after each driving cycle is completed. We can quantify the extent to which the Floquet condition is satisfied by calculating the determinant of the overlap matrix $S_{ij} = \langle \psi_i(0)|\psi_j(T)\rangle$, between the initial time-dependent Kohn-Sham (TD-KS) orbitals and those after one driving cycle has finished, as shown in Figure 1, and also discussed in Ref. [22]. When the Floquet condition (i.e. |det(**S**)|=1) is satisfied, the integrated current, $Q(T)$, is equal to the winding number, $W$, an integer[22, 24]. The topological phase is identified as having a non-zero integer value of $W$ while the $W=0$ phase indicates the trivial insulator phase.

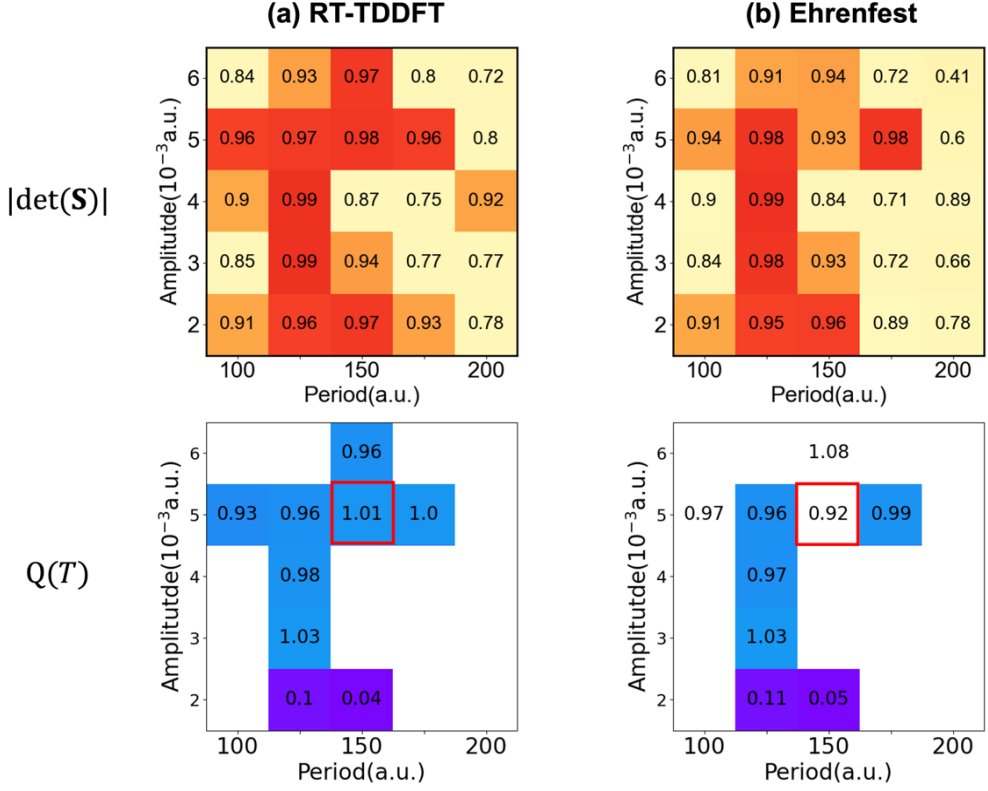

**Figure 1:** (Top) The determinant of the overlap matrix, **S**, between the initial and final TD-KS states in a single driving cycle and (Bottom) the time-integrated current over a single driving cycle $Q(T)$ plotted as a function of the driving electric field amplitude |**A**| and the period $T$ for (a) RT-TDDFT and (b) Ehrenfest dynamics. the colors indicate the value of |det(**S**)| in Top figures; red (>0.95), orange (0.90~0.95), and yellow (<0.90). In Bottom figures, the purple and blue colors indicate that the assignment of the trivial and topological phases, respectively. The red box indicates the specific driving field for which the electron dynamics were analysed in detail (see the main text).

In RT-TDDFT simulation, the lattice (atomic nuclei) is held fixed in time at the equilibrium positions. In Ehrenfest dynamics simulation, the atomic nuclei are allowed to respond to the time-dependent movement of electrons in the system according to the instantaneous Coulomb force on the atomic nuclei. The atomic nuclei thus move and respond to the non-equilibrium electron density and to the external driving field. Figure 1 (Top) shows the extent to which the Floquet condition is satisfied; the red-colored areas (indicating |det(**S**)| > 0.95) decrease when the dynamical coupling between the electrons and atomic nuclei is taken into account as in the Ehrenfest dynamics simulation. As perhaps expected, the higher the driving field amplitude is the more differences we tend to observe between RT-TDDFT and Ehrenfest dynamic simulations. Figure 1 (Bottom) shows $Q(T)$ per monomer as a function of the driving field amplitude and period. As can be seen, the Floquet topological phase remains largely intact even with the dynamical coupling between electrons and atomic nuclei. For the driving field amplitude above $|A|=2 \times 10^{-3}$ a.u. (1 a.u. ≈ 5.1422 V/Å), RT-TDDFT simulation shows that the topological phase is observed when the Floquet condition is satisfied. With Ehrenfest dynamics simulation, the lattice movement somewhat makes it more difficult to satisfy the Floquet condition; a notable change is observed for $T$=150 a.u./$|A|=5 \times 10^{-3}$ a.u.. (as marked by the red square box in Fig. 1(Bottom)). Floquet topological phase ($W$=1) can be found here as opposed to the trivial phase in the RT-TDDFT simulation. With the dynamical coupling of electrons and atomic nuclei in Ehrenfest dynamics, |Det(S)| is only 0.93, and Floquet condition is not lifted appreciably. The integrated current gives $Q$(T)=0.92, and it deviates noticeably from the integer value of $Q$(T)=1 as one would expect for the topological phase even if one considers errors from numerical simulations. The different becomes more evident when $Q(t)$ is plotted as a function of time as done in Figure 2 (a). This quantity can be conveniently calculated using the Wannier centers as in Eq. (1) from the RT-TDDFT and Ehrenfest Dynamics simulations. Both simulations show identical behavior up until the last $t$=25 a.u.. However, for the Ehrenfest dynamics, $Q$(t) value starts to decrease, indicating that the transport direction of the C-C and C=C Wannier functions reverses to the opposite direction at $t$ =125~150 a.u.. Then, the time-dependent electron density does not evolve back to the initial state after a single driving cycle, leaving some electrons excited after the periodic time, T (see Figure 2(b)); the Floquet condition is not satisfied because the single-particle KS Hamiltonian is dependent of the time-dependent density.

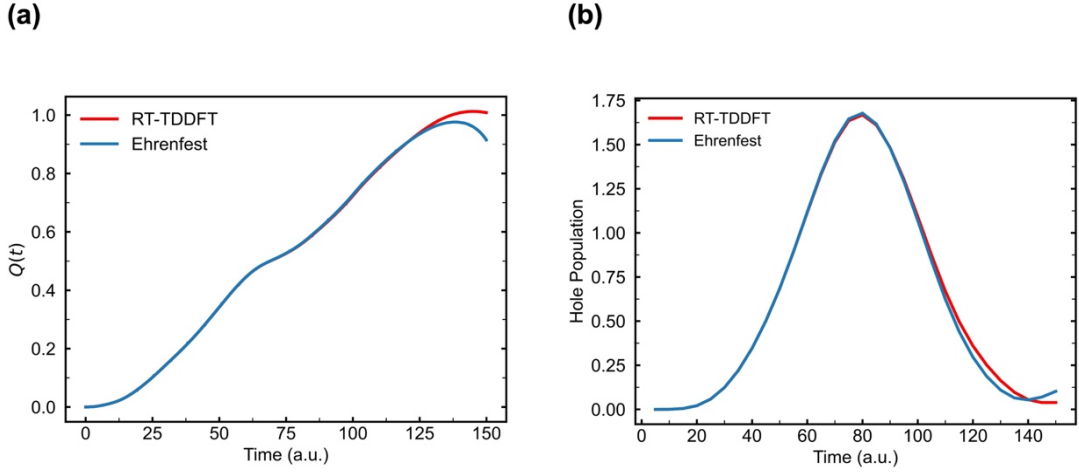

**Figure 2:** Electron dynamics from RT-TDDFT and Ehrenfest dynamics simulation of trans-polyacetylene with the driving field of $T$=150 a.u./$|A|$=5 × $10^{-3}$ a.u. (as marked by the red box in Figure 1). (a) Time-integrated current $Q(t)$ as a function of time $t$. (b) Time-dependent hole occupation, calculated by projecting the time-dependent orbitals onto the equilibrium Kohn-Sham eigenstates.

**Thermal fluctuation of lattice on electron transport**

Let us now discuss the impact of thermal fluctuations of atoms at room temperature on the Floquet topological phase. In order to sample representative geometries, first-principal molecular dynamics (FPMD) simulation was performed in the electronic ground state at 300K for 4 picoseconds. FPMD trajectories show fluctuations in both C-C and C=C bond lengths. At the same time, the analysis of the Wannier functions shows that the alternating bonding pattern, resulting from the Peierls instability, is consistently observed throughout the FPMD simulation. To quantify the extent of the Peierls distortion effect[28], we employ a collective variable, called bond length alternation (BLA)[58] index. The BLA is defined as the averaged bond length differences between the C-C and C=C bonds

$$BLA = \frac{1}{N} \sum_{i=1}^{N} (r_i^{C=C} - r_i^{C-C})$$

where $N$ represents the number of the unit cells in the simulation supercell. $r_i^{C=C}$ and $r_i^{C-C}$ are the lengths of the carbon-carbon double bond and carbon-carbon single bond in the $i$-th unit cell, respectively. We considered only the absolute value of BLA due to the inversion symmetry of trans-polyacetylene although BLA can take negative values[58]. BLA is a geometric measure of conjugated polymers, and other studies have shown that various electronic[58-60] and optical[61-62] properties can be rationalized in terms of BLA value. To better establish the relationship between the BLA index and Peierls distortion in trans-polyacetylene at the electronic structure level, the maximally localized Wannier functions (MLWF) are calculated on several selected structures from the FPMD trajectory. Figure 3 shows how the average spread values of the MLWFs associated with C=C double bonds are correlated with the BLA value. The dashed vertical line indicates the equilibrium structure at 0 K. C=C bonds tend to become, on average, more delocalized with the decreasing BLA value while C-C bonds remain largely unaffected. While the average MLWF spreads show a rather large standard deviation for the double bonds, as indicated by the error bars in Fig. 3, a clear distinction between the double and single bonds can be made for all structures with various BLA values, indicating the electronic conjugation remains intact even at the room temperature.

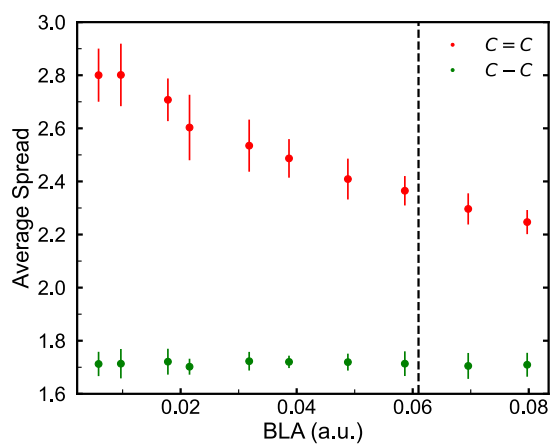

**Figure 3:** Average spread values of C=C (red) and C-C (green) Maximally-localized Wannier functions (MLWFs) in the electronic ground state for structures with different Bond Length Alternation (BLA) values. The structures are taken from the trajectory of FPMD simulation of trans-polyacetylene at 300K. The vertical dashed line represents the BLA value of the equilibrium (0K) structure. The error bars indicate the standard deviations among the MLWFs for each bond type.

RT-TDDFT calculation was performed on four selected structures with different BLA values from the FPMD simulation, as shown in Figure 4. The Floquet topological phase can be observed in all four cases with the driving field period of approximately 100-150 a.u., which aligns well with our earlier findings[22, 24]. At the same time, the extent to which the topological phase is present is highly dependent on the driving field amplitude and period (see Fig. 4 Top). Due to the dynamical disruption to the periodicity of the otherwise perfectly repeating polymer units caused by thermal fluctuations (see Fig. 3), satisfying the Floquet condition with the external driving field becomes more restrictive. Even when the Floquet condition is satisfied, whether the trivial phase (W=0) or the topological phase (W=1) exists is very sensitive to the driving field condition, strongly varying with the BLA value. For instance, for Structure A with the small BLA value of 0.010, not only it is more difficult to have the Floquet condition satisfied but emergence of the topological phase is highly dependent on the driving field, particularly with respect to the period of the driving field. In experiments performed at room temperature, all these structural variations with different BLA values would exist and thus finding the driving field condition under which the Floquet topological phase is observed at all times is likely difficult.

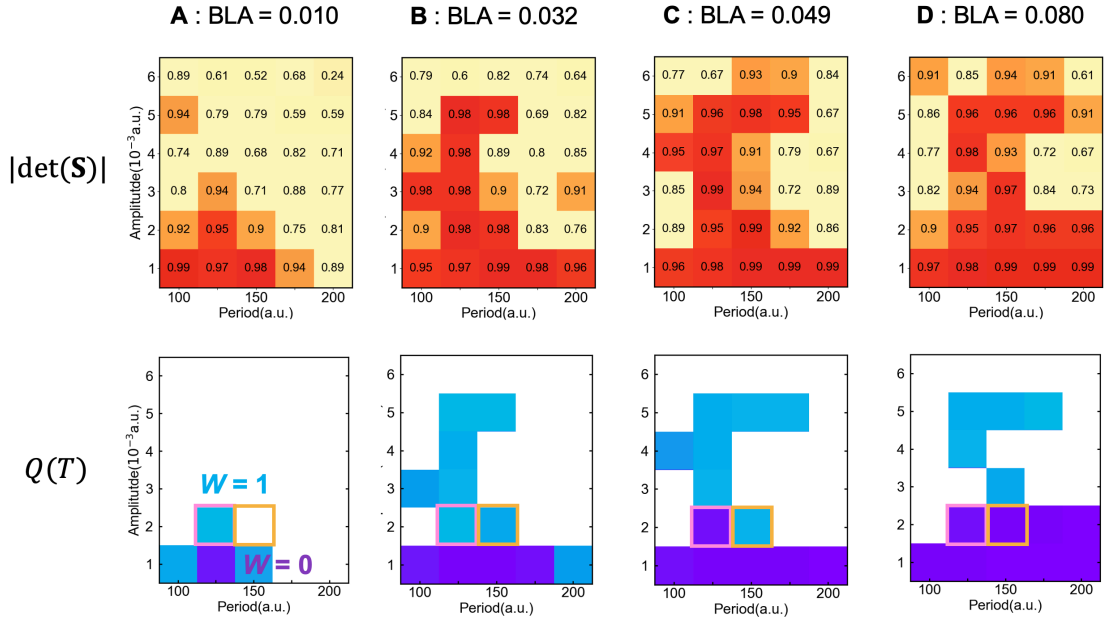

**Figure 4:** (Top) Time-integrated current over one driving cycle Q(T), and (Bottom) determinant of the overlap matrix, **S**, between the initial and final TD-KS states in a single driving cycle. These values are plotted as a function of the driving electric field amplitude |**A**| and the time period T for the structures (A, B, C, D) with various BLA values. These structures are taken from FPMD simulation of trans-polyacetylene at 300K. The white (colorless) areas in the Top plot represents that the Floquet condition is not satisfied. The pink (T = 125 a.u. and |A| = 2 × 10$^{-3}$ a.u.) and orange (T = 125 a.u. and |A| = 2 × 10$^{-3}$ a.u.) boxes indicates the specific condition for which the underlying electron dynamics were investigated in detail as discussed in the main text.

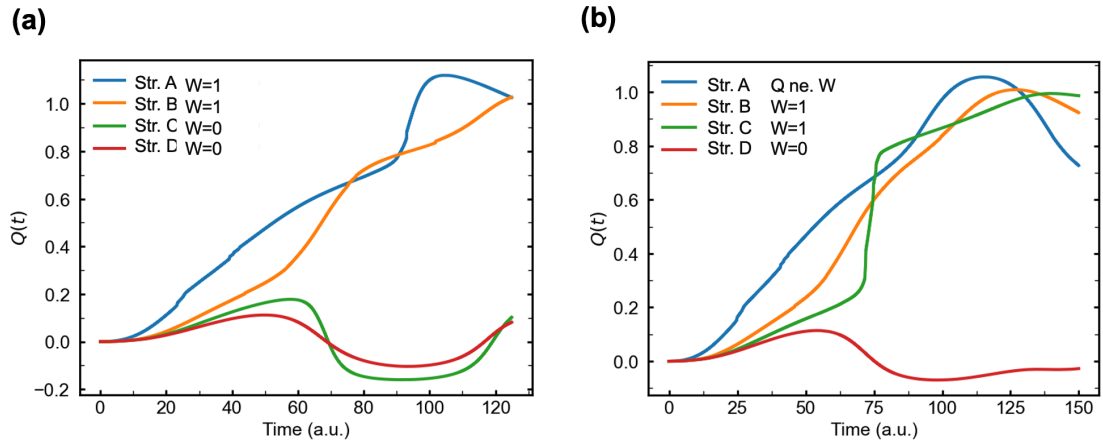

**Figure 5.** Time-integrated current as a function of time, $Q(t)$, of trans-polyacetylene for the structures of various BLA values with the driving field of (a) T=125 a.u./|A|=2 × 10$^{-3}$ a.u. (as marked by the pink square box in Fig. 4) (b) T=150 a.u./|A|=2 × 10$^{-3}$ a.u. (as marked by the orange square box in Fig. 4).

To gain a deeper insight into the quantum dynamic behavior underlying the Floquet topological pump, we discuss two specific conditions for the driving field; T = 125 a.u./|**A**| = 2 × 10$^{-3}$ a.u. (pink square in Fig. 4) and T = 150 a.u./|**A**| = 2 × 10$^{-3}$ a.u. (orange square in Fig. 4). With T = 125 a.u./|**A**| = 2 × 10$^{-3}$ a.u., the Floquet topological phase is observed for Structures A (BLA=0.010) and B (BLA=0.032) while the trivial phase (i.e. W=0) is found for Structures C (BLA=0.049) and D (BLA=0.080) as shown in Figure 4. Figure 5 show the number of transported charges $Q(t)$ as a function of time for all these driving conditions for the four structures. With the driving field of T = 125 a.u./|**A**| = 2 × 10$^{-3}$ a.u. (Fig. 5 (a)), both Structures A and B yield $Q(T)$=1, exhibiting the topological phase although their corresponding quantum dynamics show notable differences. In Structure B, $Q(t)$ shows a continuous increase, reaching a value of 1.026 electrons in the end of single driving cycle. In contrast, for Structure A, the current reverses its direction at around t=110 a.u. such that $Q(t)$ decreases before reaching the value of 1.028 electrons at the end of the driving cycle. For Structures C and D, the current direction more or less follows the direction of the driving electric field and thus $Q(t)$ is diminishingly small after the period T.

With the driving field of T = 150 a.u./|**A**| = 2 × 10$^{-3}$ a.u. (Fig. 5 (b)), the Floquet topological phase is observed for Structures B and C while the trivial phase is found for Structure D. Even accounting for some numerical errors in simulation, the Floquet condition is not satisfied for Structure A and $Q(T)$ is 0.73 electrons. Unlike for Structure A, Structures B and C both largely satisfy the Floquet condition given the same criterion based on the |det(**S**)|. While they tend to exhibit the Floquet topological phase instead of the trivial phase, the degree to which Q(T) value is close to the integer value of one is noticeably less for Structure C; $Q(T)$ for Structure B and Structure C is 0.99 and 0.92, respectively (see Supporting Information). For these structures taken from the room temperature FPMD simulation, the periodicity is not perfectly satisfied and noticeable deviations of $Q(T)$ from $Q(T)$=W=1 can be anticipated, as discussed earlier. In Figure 5, another notable feature is the abrupt jump in $Q(t)$ for Structure C around t=75 a.u.,

and such quantum tunnelling-like behavior has been discussed previously[23, 52]. If computational resources were not a factor in these highly demanding RT-TDDFT simulations, one would ideally calculate the ensemble average of $Q(t)$ to model experiments at room temperature. Figure 5, even with only four representative geometries, shows that such an ensemble-average $\langle Q(t) \rangle$ would not yield the topological phase behavior at 0K (i.e. an integer value at t=T) even for the most likely/favorable driving field condition (T = 125 a.u./|**A**| = 2 × 10$^{-3}$ a.u. (Fig. 5 (a)).

Lastly, the results here show that Bond Length Alternation (BLA) is an effective geometric parameter for the emergence and characteristics of the Floquet topological phase. As BLA value decreases, weakening Peierls distortion leads to electron delocalization within the conjugated chain. In the delocalization limit with BLA=0, the system becomes metallic with no Peierls distortions. As conveniently seen using Su-Schrieffer-Heeger model of Peierls instability[63], a topological phase is precluded in such a gapless metallic system even with an external driving field. At elevated temperatures, thermal fluctuations are more likely to dynamically disrupt the delicate electronic ordering that is required for the topological phase since the geometries with low BLA values are more accessible.

**Conclusion**

Building on the previous work[22, 24], we examined the robustness of the Floquet topological phase for trans-polyacetylene with respect to the atomic lattice movement using first-principles simulation. In particular, the effect of atomic lattice dynamics on the Floquet topological phase was investigated by performing the Ehrenfest dynamics simulation in which atoms respond to the electronic current. Secondly, we investigated how atomic lattice distortion at room temperature could affect the presence of the Floquet topological phase by performing RT-TDDFT calculations on selected geometries from a separate FPMD simulation at room temperature. When the dynamical coupling between electrons and atomic nuclei is included as in Ehrenfest dynamics simulation, it becomes somewhat more difficult to satisfy the Floquet condition although the topological phase is largely intact. Additionally, we examined the impact of thermal fluctuation in the atomic lattice on the Floquet topological phase. Bond Length Alternation (BLA) index was used to quantify the degree of Peierls distortion in terms of the carbon-carbon single and double bonds. The sensitivity of the topological phase to the driving field varies substantially with BLA value. Therefore, finding the necessary driving field magnitude and period for inducing the Floquet topological phase would be challenging at room temperature in experiments. In this theoretical work, the nuclear quantum fluctuation, as often represented by Wigner distribution, on the Floquet topological phase was not considered. Such a nuclear quantum effect could introduce additional challenges to observing the Floquet topological phase even at low temperatures, and it will be studied in future studies. Lastly, the exchange-correlation (XC) approximation in first-principles simulation was found to have a great impact on the prediction of the Peierls distortion for trans-polyacetylene. Improving XC approximation continues to be of great practical importance and such efforts continue to be necessary in the community.

**Acknowledgment**

This work was supported by the National Science Foundation under Award Nos. CHE-1954894 and OAC-2209858. We thank the Research Computing at the University of North Carolina at Chapel Hill for providing computational resources.**Data Availability Statement**

The data supporting this article have been included as part of the Supplementary Information.

**Reference**
1. Thouless, D. J., Quantization of Particle Transport. *Physical Review B* **1983**, *27*, 6083-6087.
2. Ma, W.; Zhou, L.; Zhang, Q.; Li, M.; Cheng, C.; Geng, J.; Rong, X.; Shi, F.; Gong, J.; Du, J., Experimental Observation of a Generalized Thouless Pump with a Single Spin. *Physical review letters* **2018**, *120*, 120501.
3. Cerjan, A.; Wang, M.; Huang, S.; Chen, K. P.; Rechtsman, M. C., Thouless Pumping in Disordered Photonic Systems. *Light: Science & Applications* **2020**, *9*, 1-7.
4. Guo, A.-M.; Hu, P.-J.; Gao, X.-H.; Fang, T.-F.; Sun, Q.-F., Topological Phase Transitions of Thouless Charge Pumping Realized in Helical Organic Molecules with Long-Range Hopping. *Physical Review B* **2020**, *102*, 155402.
5. Zhang, Y.; Gao, Y.; Xiao, D., Topological Charge Pumping in Twisted Bilayer Graphene. *Physical Review B* **2020**, *101*, 041410.
6. Nakajima, S.; Tomita, T.; Taie, S.; Ichinose, T.; Ozawa, H.; Wang, L.; Troyer, M.; Takahashi, Y., Topological Thouless Pumping of Ultracold fermions. *Nature Physics* **2016**, *12*, 296.
7. Lohse, M.; Schweizer, C.; Zilberberg, O.; Aidelsburger, M.; Bloch, I., A Thouless Quantum Pump with Ultracold Bosonic Atoms in an Optical Superlattice. *Nature Physics* **2016**, *12*, 350-354.
8. Hayward, A.; Schweizer, C.; Lohse, M.; Aidelsburger, M.; Heidrich-Meisner, F., Topological Charge Pumping in the Interacting Bosonic Rice-Mele Model. *Physical Review B* **2018**, *98*, 245148.
9. Friedman, A. J.; Gopalakrishnan, S.; Vasseur, R., Integrable Many-Body Quantum Floquet-Thouless Pumps. *Physical review letters* **2019**, *123*, 170603.
10. Taherinejad, M.; Garrity, K. F.; Vanderbilt, D., Wannier Center Sheets in Topological Insulators. *Physical Review B* **2014**, *89*, 115102.
11. Li, R.; Fleischhauer, M., Finite-Size Corrections to Quantized Particle Transport in Topological Charge Pumps. *Physical Review B* **2017**, *96*, 085444.
12. Rice, M. J.; Mele, E. J., Elementary Excitations of a Linearly Conjugated Diatomic Polymer. *Physical Review Letters* **1982**, *49*, 1455-1459.
13. Privitera, L.; Russomanno, A.; Citro, R.; Santoro, G. E., Nonadiabatic Breaking of Topological Pumping. *Physical Review Letters* **2018**, *120*, 106601.
14. Kuno, Y., Non-Adiabatic Extension of the Zak Phase and Charge Pumping in the Rice–Mele Model. *The European Physical Journal B* **2019**, *92*, 195.
15. Fedorova, Z.; Qiu, H.; Linden, S.; Kroha, J. In *Topological Transport Quantization by Dissipation in Fast Thouless Pumps*, CLEO: QELS_Fundamental Science, Optical Society of America: 2020; p FM2A. 3.
16. Oka, T.; Kitamura, S., Floquet Engineering of Quantum Materials. *Annual Review of Condensed Matter Physics* **2019**, *10*, 387-408.
17. Rudner, M. S.; Lindner, N. H., Band Structure Engineering and Non-Equilibrium Dynamics in Floquet Topological Insulators. *Nature Reviews Physics* **2020**, *2*, 229-244.

18. Khoury, J. F.; Schoop, L. M., Chemical Bonds in Topological Materials. *Trends in Chemistry* **2021**, *3*, 700-715.
19. Martín Pendás, A.; Contreras-García, J.; Pinilla, F.; Mella, J. D.; Cardenas, C.; Muñoz, F., A Chemical Theory of Topological Insulators. *Chemical Communications* **2019**, *55*, 12281-12287.
20. Castro, A.; De Giovannini, U.; Sato, S. A.; Huebener, H.; Rubio, A., Floquet Engineering with Quantum Optimal Control Theory. *New Journal of Physics* **2023**.
21. Topp, G. E.; Jotzu, G.; McIver, J. W.; Xian, L.; Rubio, A.; Sentef, M. A., Topological Floquet Engineering of Twisted Bilayer Graphene. *Physical Review Research* **2019**, *1*, 023031.
22. Zhou, R.; Yost, D. C.; Kanai, Y., First-Principles Demonstration of Nonadiabatic Thouless Pumping of Electrons in a Molecular System. *The Journal of Physical Chemistry Letters* **2021**, *12*, 4496-4503.
23. Zhou, R.; Kanai, Y., Dynamical Transition Orbitals: A Particle–Hole Description in Real-Time Tddft Dynamics. *The Journal of Chemical Physics* **2021**, *154*, 054107.
24. Zhou, R.; Kanai, Y., Molecular Control of Floquet Topological Phase in Non-Adiabatic Thouless Pumping. *The Journal of Physical Chemistry Letters* **2023**, *14*, 8205-8212.
25. Perdew, J. P.; Burke, K.; Ernzerhof, M., Generalized Gradient Approximation Made Simple. *Physical Review Letters* **1996**, *77*, 3865-3868.
26. Sun, J.; Ruzsinszky, A.; Perdew, J. P., Strongly Constrained and Appropriately Normed Semilocal Density Functional. *Physical review letters* **2015**, *115*, 036402.
27. Adamo, C.; Barone, V., Toward Reliable Density Functional Methods without Adjustable Parameters: The Pbe0 Model. *The Journal of chemical physics* **1999**, *110*, 6158-6170.
28. Ashkenazi, J.; Pickett, W.; Krakauer, H.; Wang, C.; Klein, B.; Chubb, S., Ground State of Trans-Polyacetylene and the Peierls Mechanism. *Physical review letters* **1989**, *62*, 2016.
29. Hui, K.; Chai, J.-D., Scan-Based Hybrid and Double-Hybrid Density Functionals from Models without Fitted Parameters. *The Journal of chemical physics* **2016**, *144*.
30. Yannoni, C.; Clarke, T., Molecular Geometry of Cis-and Trans-Polyacetylene by Nutation Nmr Spectroscopy. *Physical review letters* **1983**, *51*, 1191.
31. Ciofini, I.; Adamo, C.; Chermette, H., Effect of Self-Interaction Error in the Evaluation of the Bond Length Alternation in Trans-Polyacetylene Using Density-Functional Theory. *The Journal of chemical physics* **2005**, *123*.
32. Blum, V.; Rossi, M.; Kokott, S.; Scheffler, M., The Fhi-Aims Code: All-Electron, Ab Initio Materials Simulations Towards the Exascale. *arXiv preprint arXiv:2208.12335* **2022**.
33. Blum, V.; Gehrke, R.; Hanke, F.; Havu, P.; Havu, V.; Ren, X.; Reuter, K.; Scheffler, M., Ab Initio Molecular Simulations with Numeric Atom-Centered Orbitals. *Computer Physics Communications* **2009**, *180*, 2175-2196.
34. Nosé, S., A Unified Formulation of the Constant Temperature Molecular Dynamics Methods. *The Journal of chemical physics* **1984**, *81*, 511-519.

35. Hoover, W. G., Canonical Dynamics: Equilibrium Phase-Space Distributions. *Physical review A* **1985**, *31*, 1695.
36. Shirakawa, H.; Ikeda, S., Infrared Spectra of Poly (Acetylene). *Polymer Journal* **1971**, *2*, 231-244.
37. Runge, E.; Gross, E. K. U., Density-Functional Theory for Time-Dependent Systems. *Physical Review Letters* **1984**, *52*, 997-1000.
38. Yabana, K.; Bertsch, G. F., Time-Dependent Local-Density Approximation in Real Time. *Physical Review B* **1996**, *54*, 4484-4487.
39. Xu, J.; Carney, T. E.; Zhou, R.; Shepard, C.; Kanai, Y., Real-Time Time-Dependent Density Functional Theory for Simulating Nonequilibrium Electron Dynamics. *Journal of the American Chemical Society* **2024**, *146*, 5011-5029.
40. Horsfield, A.; Finnis, M.; Foulkes, M.; LePage, J.; Mason, D.; Race, C.; Sutton, A.; Bowler, D.; Fisher, A.; Miranda, R., Correlated Electron-Ion Dynamics in Metallic Systems. *Computational Materials Science* **2008**, *44*, 16-20.
41. Loaiza, I.; Izmaylov, A. F., On the Breakdown of the Ehrenfest Method for Molecular Dynamics on Surfaces. *The Journal of Chemical Physics* **2018**, *149*.
42. Schleife, A.; Draeger, E. W.; Anisimov, V. M.; Correa, A. A.; Kanai, Y., Quantum Dynamics Simulation of Electrons in Materials on High-Performance Computers. *Computing in Science & Engineering* **2014**, *16*, 54-60.
43. Draeger, E.; Gygi, F., Qbox Code, Qb@ Ll Version. *Lawrence Livermore National Laboratory* **2017**.
44. Shepard, C.; Zhou, R.; Yost, D. C.; Yao, Y.; Kanai, Y., Simulating Electronic Excitation and Dynamics with Real-Time Propagation Approach to Tddft within Plane-Wave Pseudopotential Formulation. *The Journal of Chemical Physics* **2021**, *155*, 100901.
45. Kononov, A.; Lee, C.-W.; dos Santos, T. P.; Robinson, B.; Yao, Y.; Yao, Y.; Andrade, X.; Baczewski, A. D.; Constantinescu, E.; Correa, A. A., Electron Dynamics in Extended Systems within Real-Time Time-Dependent Density-Functional Theory. *MRS Communications* **2022**, *12*, 1002-1014.
46. Shepard, C.; Zhou, R.; Bost, J.; Carney, T. E.; Yao, Y.; Kanai, Y., Efficient Exact Exchange Using Wannier Functions and Other Related Developments in Planewave-Pseudopotential Implementation of Rt-Tddft. *The Journal of Chemical Physics* **2024**, *161*.
47. Francois, G., Architecture of Qbox: A Scalable First-Principles Molecular Dynamics Code. *IBM Journal of Research and Development* **2008**, *52*, 137-144.
48. Schleife, A.; Draeger, E. W.; Kanai, Y.; Correa, A. A., Plane-Wave Pseudopotential Implementation of Explicit Integrators for Time-Dependent Kohn-Sham Equations in Large-Scale Simulations. *The Journal of Chemical Physics* **2012**, *137*, 22A546.
49. Hamann, D. R.; Schlüter, M.; Chiang, C., Norm-Conserving Pseudopotentials. *Physical Review Letters* **1979**, *43*, 1494-1497.
50. Vanderbilt, D., Optimally Smooth Norm-Conserving Pseudopotentials. *Physical Review B* **1985**, *32*, 8412-8415.

51. Castro, A.; Marques, M. A.; Rubio, A., Propagators for the Time-Dependent Kohn–Sham Equations. *The Journal of chemical physics* **2004**, *121*, 3425-3433.
52. Yost, D. C.; Yao, Y.; Kanai, Y., Propagation of Maximally Localized Wannier Functions in Real-Time Tddft. *The Journal of chemical physics* **2019**, *150*, 194113.
53. Resta, R., Quantum-Mechanical Position Operator in Extended Systems. *Physical Review Letters* **1998**, *80*, 1800.
54. Nakagawa, M.; Slager, R.-J.; Higashikawa, S.; Oka, T., Wannier Representation of Floquet Topological States. *Physical Review B* **2020**, *101*, 075108.
55. Vanderbilt, D., *Berry Phases in Electronic Structure Theory: Electric Polarization, Orbital Magnetization and Topological Insulators*; Cambridge University Press: Cambridge, 2018.
56. Ullrich, C. A., *Time-Dependent Density-Functional Theory, Concepts and Applications*; Oxford Graduate Texts, 2011.
57. Marques, M. A.; Gross, E. K., Time-Dependent Density Functional Theory. *Annu. Rev. Phys. Chem.* **2004**, *55*, 427-455.
58. Bredas, J.-L., Relationship between Band Gap and Bond Length Alternation in Organic Conjugated Polymers. *The Journal of chemical physics* **1985**, *82*, 3808-3811.
59. Jacquemin, D.; Adamo, C., Bond Length Alternation of Conjugated Oligomers: Wave Function and Dft Benchmarks. *Journal of chemical theory and computation* **2011**, *7*, 369-376.
60. Yang, S.; Kertesz, M., Bond Length Alternation and Energy Band Gap of Polyyne. *The Journal of Physical Chemistry A* **2006**, *110*, 9771-9774.
61. Gieseking, R. L.; Risko, C.; Bredas, J.-L., Distinguishing the Effects of Bond-Length Alternation Versus Bond-Order Alternation on the Nonlinear Optical Properties of Π-Conjugated Chromophores. *The journal of physical chemistry letters* **2015**, *6*, 2158-2162.
62. Jacquemin, D.; Femenias, A.; Chermette, H.; Ciofini, I.; Adamo, C.; André, J.-M.; Perpète, E. A., Assessment of Several Hybrid Dft Functionals for the Evaluation of Bond Length Alternation of Increasingly Long Oligomers. *The Journal of Physical Chemistry A* **2006**, *110*, 5952-5959.
63. Asbóth, J. K.; Oroszlány, L.; Pályi, A.; Asbóth, J. K.; Oroszlány, L.; Pályi, A., The Su-Schrieffer-Heeger (Ssh) Model. *A Short Course on Topological Insulators: Band Structure and Edge States in One and Two Dimensions* **2016**, 1-22.